\documentclass[twocolumn]{revtex4}
\usepackage{amsfonts}
\usepackage{amsmath}
\usepackage{amssymb}
\usepackage{charter}
\usepackage{graphicx}

\begin{document}

\title{Conditional generation of multiphoton-subtracted squeezed vacuum
states: loss consideration and operator description}
\author{Xue-xiang Xu$^{1,\dag }$ and Hong-chun Yuan$^{2}$}
\affiliation{$^{1}$College of Physics and Communication Electronics, Jiangxi Normal
University, Nanchang 330022, China;\\
$^{2}$College of Electrical and Information Engineering, Changzhou Institute
of Technology, Changzhou 213032, China\\
$^{\dag }$E-mail: xuxuexiang@jxnu.edu.cn}

\begin{abstract}
In terms of the characteristic functions of the quantum states, we present a
complete operator description of a lossy photon-subtraction scheme. Feeding
a single-mode squeezed vacuum into a variable beam splitter and counting the
photons in one of the output channels, a broad class of
multiphoton-subtracted squeezed vacuum states (MSSVSs) can be generated in
other channel. Here the losses are considered in the beginning and the end
channels in the circuit. Indeed, this scheme has been discussed in Ref.
[Phys. Rev. A 100, 022341 (2019)]. However, different from the above work,
we give all the details of the optical fields in all stages. In addition, we
present the analytical expressions and numerical simulations for the success
probability, the quadrature squeezing effect, photon-number distribution and
Wigner function of the MSSVSs. Some interesting results effected by the
losses are obtained.


\end{abstract}

\maketitle

\section{Introduction}

A key requirement for many quantum protocols is to use specific quantum
states of light as a resource for information processing\cite{1}. These
quantum states can be divided into Gaussian and non-Gaussian cases\cite{2}.
For example, the coherent state and the single-mode squeezed vacuum state
are the typical Gaussian state, which are applied in many tasks\cite{3}.
However, various important protocols for quantum enhanced information
processing cannot be performed when restricted to Gaussian states\cite{4}.
Thus, it is necessary to introducing non-Gaussianity into an optical system.
In recent years, many non-Gaussian quantum states have been used as
resources for useful quantum information processing tasks \cite{5,6}.
Therefore, a crucial goal for experimental quantum optics is to prepare
high-quality non-Gaussian quantum states.

Photon-subtraction operation is just a useful way to conditionally
manipulate a non-Gaussian state of the optical field, which has been shown
to enhance entanglement\cite{7,8} and teleportation fidelity\cite{9}.
Theoretically, subtraction of $m$ photons from a single-mode quantum state $%
\left\vert \psi \right\rangle $\ can be expressed as $a^{m}\left\vert \psi
\right\rangle $, where $a$\ is the photon annihilation operator\cite{9-1}.
Experimentally, such photon-subtracted state can be implemented by
transmitting $\left\vert \psi \right\rangle $\ through a beam splitter and
detecting the output of the beam splitter with photon number resolving
detector\cite{9-2}. Studies have shown that photon subtraction on a
single-mode squeezed vacuum state yields optical coherent-state superposition%
\cite{10} or Schrodinger-cat-like states\cite{11}.

As a matter of fact, the loss is unavoidable in the propagating channel of
light beams. It is necessary to analyze and control the effect of loss in
the quantum protocols\cite{12,13}. Very recently, Quesada et al. considered
some\ schemes of preparing conditional non-Gaussian states in the presence
of photon loss. Among them, the photon-subtraction scheme is more attracted
our attention\cite{14}. In the present paper, we shall give a description
for the same scheme in terms of the characteristic function (CF) of the
quantum states involved. The CF of density operator $\rho $ can be defined
as $\chi _{\rho }\left( \alpha \right) =\mathrm{Tr}[\rho D\left( \alpha
\right) ]$, i.e. the expectation value of the Weyl displacement operator $%
D\left( \alpha \right) =\exp \left( \alpha a^{\dag }-\alpha ^{\ast }a\right)
$ \cite{15}. One main tool in dealing with optical field is the Weyl
expansion of the density operator, that is,%
\begin{equation}
\rho =\int \frac{d^{2}\alpha }{\pi }\chi _{\rho }(\alpha )D\left( -\alpha
\right) ,  \label{a1}
\end{equation}%
which means that the function $\chi _{\rho }(\alpha )$\ uniquely determines
the density operator $\rho $ \cite{16,17}. Another tool in deriving
input-output relation of loss channel $L\left( \eta \right) $ (with loss
factor $\eta \in \lbrack 0,1]$) is that the output density operator $\rho
_{out}$ can be expressed as the integration form of the input CF $\chi
_{\rho _{in}}\left( \alpha \right) $, i.e.%
\begin{equation}
\rho _{out}=\int \frac{d^{2}\alpha }{\pi }\chi _{\rho _{in}}(\sqrt{1-\eta }%
\alpha )D\left( -\alpha \right) e^{-\frac{1}{2}\eta \left\vert \alpha
\right\vert ^{2}},  \label{a2}
\end{equation}%
which describes that $\rho _{in}$ evolutes into $\rho _{out}$ through loss
channel. This equation has been derived in our previous work\cite{18}.

The paper is organized as follows. In Sec.2, we introduce the density
operator description of generating such state. Here we shall give the
conceptual scheme and decompose the whole circuit into five stages, whose
density operators are derived. Then in Sec. 3-5, we study the analytical and
numerical results of the properties related with our generated states,
including quadrature squeezing effect, photon number distribution and\
Wigner function, and explore how the photon subtraction and loss factors
affect these nonclassicalities. Finally, a summary is given in Sec.6.

\section{Density operator description of theoretical scheme}

Fig.1 shows the conceptual scheme of generating multiphoton-subtracted
squeezed vacuum states (MSSVSs). The whole circuit can be decomposed into
five stages, that is%
\begin{equation}
\rho _{ab}^{\left( 1\right) }\overset{L\left( \eta _{1}\right) }{\implies }%
\rho _{ab}^{\left( 2\right) }\overset{B\left( \theta \right) }{\implies }%
\rho _{ab}^{\left( 3\right) }\overset{L\left( \eta _{2}\right) }{\implies }%
\rho _{ab}^{\left( 4\right) }\overset{\left\vert m\right\rangle \left\langle
m\right\vert }{\implies }\rho .  \label{a0}
\end{equation}%
In this scheme, red line and blue line are corresponding to mode a and mode
b, respectively. $\rho _{ab}^{\left( 1\right) }$ is the direct product of
the single-mode squeezed vacuum state (SVS) $S\left( r\right) \left\vert
0\right\rangle $ and the vacuum $\left\vert 0\right\rangle $. Here $S\left(
r\right) =\exp [\frac{r}{2}(a^{\dag 2}-a^{2})]$ is the single-mode squeezed
operator with the real squeezing parameter $r$ \cite{19}. After $\rho
_{ab}^{\left( 1\right) }$ passing through channel $L\left( \eta _{1}\right) $
with loss factor $\eta _{1}$, $\rho _{ab}^{\left( 2\right) }$ is obtained.
Injecting $\rho _{ab}^{\left( 2\right) }$\ into a variable beam-splitter $%
B\left( \theta \right) $, $\rho _{ab}^{\left( 3\right) }$ is obtained. Then
after $\rho _{ab}^{\left( 3\right) }$\ passing through channel $L\left( \eta
_{2}\right) $ with loss factor $\eta _{2}$, $\rho _{ab}^{\left( 4\right) }$
is obtained. In the last stage, the MSSVS $\rho $\ is generated heraldedly
by performing a $m$-photon detection.
\begin{figure}[tbp]
\label{Fig1} \centering\includegraphics[width=1.0\columnwidth]{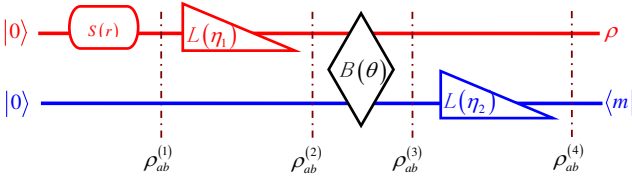}
\caption{(Colour online) Conceputal scheme of generating MPSSVs.}
\end{figure}

\textit{State in stage 1: }The total density operator in this stage can be
written as
\begin{equation}
\rho _{ab}^{\left( 1\right) }=\left( \rho _{SV}\right) _{a}\otimes \left(
\left\vert 0\right\rangle \left\langle 0\right\vert \right) _{b}  \label{a4}
\end{equation}%
with $\rho _{SV}=S\left( r\right) \left\vert 0\right\rangle \left\langle
0\right\vert S^{\dag }\left( r\right) $, where the SVS can be further
expressed as $S\left( r\right) \left\vert 0\right\rangle =\left( 1-\lambda
^{2}\right) ^{1/4}e^{\frac{\lambda }{2}a^{\dag 2}}\left\vert 0\right\rangle $
with $\lambda =\tanh r$. Therefore, $\rho _{ab}^{\left( 1\right) }$ can be
given by%
\begin{equation}
\rho _{ab}^{\left( 1\right) }=\int \frac{d^{2}\alpha _{1}d^{2}\beta _{1}}{%
\pi ^{2}}\chi _{\rho _{ab}^{\left( 1\right) }}(\alpha _{1},\beta
_{1})D_{a}\left( -\alpha _{1}\right) D_{b}\left( -\beta _{1}\right) ,
\label{a5}
\end{equation}%
where $\chi _{\rho _{ab}^{\left( 1\right) }}(\alpha _{1},\beta _{1})$ is the
CF of $\rho _{ab}^{\left( 1\right) }$ satisfying%
\begin{eqnarray}
\chi _{\rho _{ab}^{\left( 1\right) }}(\alpha _{1},\beta _{1}) &=&\mathrm{Tr}%
[\rho _{ab}^{\left( 1\right) }D_{a}\left( \alpha _{1}\right) D_{b}\left(
\beta _{1}\right) ]  \notag \\
&=&e^{-\allowbreak \frac{(1+\lambda ^{2})\left\vert \alpha _{1}\right\vert
^{2}}{2(1-\lambda ^{2})}+\frac{\lambda \allowbreak (\alpha _{1}^{2}+\alpha
_{1}^{\ast }{}^{2})}{2(1-\lambda ^{2})}-\frac{\left\vert \beta
_{1}\right\vert ^{2}}{2}},  \label{a6}
\end{eqnarray}%
and both $D_{a}\left( \alpha _{1}\right) \ $and $D_{b}\left( \beta
_{1}\right) $\ are the displacement operators in mode $a$ and mode $b$,
respectively.

\textit{State in stage 2: }After $\rho _{ab}^{\left( 1\right) }$ passing
through $L\left( \eta _{1}\right) $, we obtain
\begin{eqnarray}
\rho _{ab}^{\left( 2\right) } &=&\int \frac{d^{2}\alpha _{1}d^{2}\beta _{1}}{%
\pi ^{2}}\chi _{\rho _{ab}^{\left( 1\right) }}(\sqrt{1-\eta }\alpha
_{1},\beta _{1})  \notag \\
&&\times e^{-\frac{\eta _{1}\left\vert \alpha _{1}\right\vert ^{2}}{2}%
}D_{a}\left( -\alpha _{1}\right) D_{b}\left( -\beta _{1}\right) .  \label{a7}
\end{eqnarray}%
where we have considered the input-output formula in Eq.(\ref{a2}). It is
noted that we can still use the CF in stage 1 and it is not necessary to
calculate the CF in stage 2.

\textit{State in stage 3: }Injecting $\rho _{ab}^{\left( 2\right) }$\ into a
variable beam-splitter, we have%
\begin{equation}
\rho _{ab}^{\left( 3\right) }=B\left( \theta \right) \rho _{ab}^{\left(
2\right) }B^{\dag }\left( \theta \right) ,  \label{a10}
\end{equation}%
where the BS is described by $B\left( \theta \right) =\exp [\theta \left(
a^{\dag }b-ab^{\dag }\right) ]$ with the transmissivity $T=\cos ^{2}\theta $%
, satisfying $B\left( \theta \right) aB^{\dag }\left( \theta \right) =\sqrt{T%
}a+\sqrt{1-T}b$ and $B\left( \theta \right) bB^{\dag }\left( \theta \right)
=-\sqrt{1-T}a+\sqrt{T}b$ \cite{20}. After making the detailed derivation, $%
\rho _{ab}^{\left( 3\right) }$ can be expressed as%
\begin{eqnarray}
\rho _{ab}^{\left( 3\right) } &=&\int \frac{d^{2}\alpha _{1}d^{2}\beta _{1}}{%
\pi ^{2}}\chi _{\rho _{ab}^{\left( 1\right) }}(\sqrt{1-\eta _{1}}\alpha
_{1},\beta _{1})  \notag \\
&&\times e^{\frac{(1-\eta _{1})\left\vert \alpha _{1}\right\vert ^{2}}{2}+%
\frac{\left\vert \beta _{1}\right\vert ^{2}}{2}}  \notag \\
&&\times e^{a(\sqrt{T}\alpha _{1}^{\ast }-\allowbreak \beta _{1}^{\ast }%
\sqrt{1-T})}e^{-a^{\dag }(\sqrt{T}\alpha _{1}-\beta _{1}\sqrt{1-T})}  \notag
\\
&&\times e^{b(\alpha _{1}^{\ast }\allowbreak \sqrt{1-T}+\sqrt{T}\beta
_{1}^{\ast })}e^{-b^{\dag }(\alpha _{1}\sqrt{1-T}+\sqrt{T}\allowbreak \beta
_{1})}.  \label{a11}
\end{eqnarray}%
Since the CF of $\rho _{ab}^{\left( 3\right) }$\ is useful to calculate $%
\rho _{ab}^{\left( 4\right) }$, we must obtain its analytic expression as
follows%
\begin{eqnarray}
\chi _{\rho _{ab}^{\left( 3\right) }}(\alpha _{3},\beta _{3}) &=&\mathrm{Tr}%
[\rho _{ab}^{\left( 3\right) }D_{a}\left( \alpha _{3}\right) D_{b}\left(
\beta _{3}\right) ]  \notag \\
&=&e^{-(\frac{1}{2}+\lambda ^{2}\tau _{1})\left\vert \alpha _{3}\right\vert
^{2}+\frac{\lambda \tau _{1}}{2}(\alpha _{3}^{2}+\alpha _{3}^{\ast }{}^{2})}
\notag \\
&&\times e^{-(\frac{1}{2}+\lambda ^{2}\tau _{2})\left\vert \beta
_{3}\right\vert ^{2}+\allowbreak \frac{1}{2}\lambda \tau _{2}(\beta
_{3}^{2}+\beta _{3}^{\ast }{}^{2})}  \notag \\
&&\times e^{\lambda \tau _{3}(\alpha _{3}\beta _{3}+\alpha _{3}^{\ast }\beta
_{3}^{\ast })-\lambda ^{2}\tau _{3}(\alpha _{3}\beta _{3}^{\ast }\allowbreak
+\beta _{3}\alpha _{3}^{\ast })},  \label{a12}
\end{eqnarray}%
where%
\begin{eqnarray}
\tau _{1} &=&\left( 1-\eta _{1}\right) T/\left( 1-\lambda ^{2}\right) ,
\notag \\
\tau _{2} &=&\left( 1-\eta _{1}\right) \left( \allowbreak 1-\allowbreak
T\right) /\left( 1-\lambda ^{2}\right) , \\
\tau _{3} &=&\left( 1-\eta _{1}\right) \sqrt{T\left( 1-T\right) }/\left(
1-\lambda ^{2}\right) .  \notag
\end{eqnarray}

\textit{State in stage 4: }After $\rho _{ab}^{\left( 3\right) }$ passing
through $L\left( \eta _{2}\right) $, similarly using Eq.(\ref{a2}) we also
may obtain $\rho _{ab}^{\left( 4\right) }$,%
\begin{eqnarray}
\rho _{ab}^{\left( 4\right) } &=&\int \frac{d^{2}\alpha _{3}d^{2}\beta _{3}}{%
\pi ^{2}}\chi _{\rho _{ab}^{\left( 3\right) }}(\alpha _{3},\sqrt{1-\eta _{2}}%
\beta _{3})  \notag \\
&&\times e^{-\frac{\eta _{2}\left\vert \beta _{3}\right\vert ^{2}}{2}%
}D_{a}\left( -\alpha _{3}\right) D_{b}\left( -\beta _{3}\right) .
\label{a14}
\end{eqnarray}

\textit{State in stage 5: }At the last stage, making a $m$-photon detection,
the MPSSV\ can be obtained%
\begin{equation}
\rho =\frac{1}{p_{d}}\left\langle m\right\vert \rho _{ab}^{\left( 4\right)
}\left\vert m\right\rangle ,  \label{a15}
\end{equation}%
where $p_{d}$\ is the success probability of producing such state.
Substituting $\left\langle m\right\vert =\frac{1}{\sqrt{m!}}\frac{d^{m}}{%
d\mu ^{m}}\left\langle 0\right\vert e^{\mu b}|_{\mu =0}$ and $\left\vert
m\right\rangle =\frac{1}{\sqrt{m!}}\frac{d^{m}}{d\nu ^{m}}e^{\nu b^{\dag
}}\left\vert 0\right\rangle |_{\nu =0}$, as well as Eq.(\ref{a14}) into Eq.(%
\ref{a15}), we obtain the density operator$\allowbreak $%
\begin{eqnarray}
\rho &=&\frac{1}{m!p_{d}}\frac{d^{2m}}{d\mu ^{m}d\nu ^{m}}\exp \left( \mu
\nu \right)  \notag \\
&&\int \frac{d^{2}\alpha _{3}d^{2}\beta _{3}}{\pi ^{2}}\chi _{\rho
_{ab}^{\left( 3\right) }}(\alpha _{3},\sqrt{1-\eta _{2}}\beta _{3})  \notag
\\
&&\times e^{-\frac{(1+\eta _{2})\left\vert \beta _{3}\right\vert ^{2}}{2}%
-\mu \beta _{3}+\nu \beta _{3}^{\ast }}D_{a}\left( -\alpha _{3}\right)
|_{\mu =\nu =0},  \label{a17}
\end{eqnarray}%
and its corresponding success probability%
\begin{equation}
p_{d}=\frac{1}{m!\sqrt{\epsilon _{4}}}\frac{d^{2m}}{d\mu ^{m}d\nu ^{m}}e^{(1-%
\frac{\epsilon _{1}}{\epsilon _{4}})\mu \nu +\frac{\epsilon _{2}}{\epsilon
_{4}}\allowbreak (\mu ^{2}+\allowbreak \nu ^{2})}|_{\mu =\nu =0},
\label{a18}
\end{equation}%
with $\epsilon _{1}=\allowbreak 1+\lambda ^{2}\tau _{2}\left( 1-\eta
_{2}\right) $, $\epsilon _{2}=\frac{1}{2}\lambda \tau _{2}\left( 1-\eta
_{2}\right) $, and $\epsilon _{4}=\epsilon _{1}^{2}-4\epsilon _{2}^{2}$.
Thus, by varying all the interaction parameters, involving the input
squeezing parameter $r$, the loss factors $\eta _{1}$, $\eta _{2}$, the BS
transmissivity $T$ and the detecting photon number $m$, a broad class of
MSSVSs can be obtained.
\begin{figure}[tbp]
\label{Fig2} \centering\includegraphics[width=1.0\columnwidth]{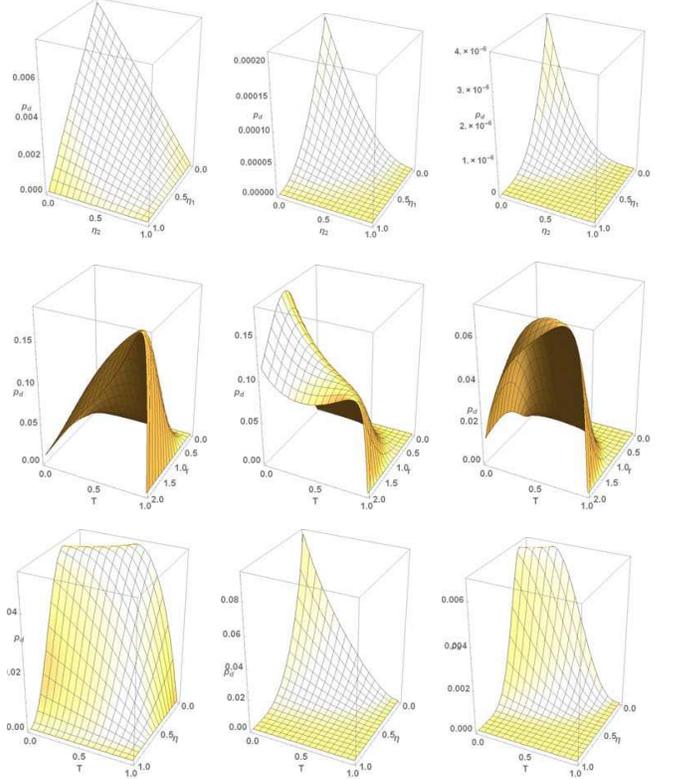}
\caption{(Colour online) Success probabilities in different paramter space
with given other parameters. (Row 1) in ($\protect\eta _{1}$, $\protect\eta %
_{2}$) space with $r=0.5$, $T=0.97$\ and different $m$; (Row 2) in ($r$, $T$%
) space with $\protect\eta _{1}=\protect\eta _{2}=0.02$; (Row 3) in ($%
\protect\eta $, $T$) space with $r=0.5$, $\protect\eta _{1}=\protect\eta %
_{2}=\protect\eta $. Columns 1, Columns 2, and Columns 3 are corresponding
to m=1, m=2, and m=3, respectively.}
\end{figure}

Success probability is an important character in the conditional quantum
state engineering. In Fig,2, according to Eq.(\ref{a18}) we plot several
distributions of success probability in different parameter spaces. We note
that the probability is limited to zero if the loss factors $\eta _{1}$, $%
\eta _{2}$ equal to 1. Next, we shall discuss the nonclassical properties of
the MSSVSs in terms of quadrature squeezing effect, photon number
distribution, Wigner function, and explore how the photon subtraction and
loss factors affect these nonclassicalities.

\section{Quadrature squeezing effect}

No doubt, the prominent character of the SVS is the quadrature squeezing
effect. But compared with the original SVS, how the squeezing effect for the
MSSVSs change? Now we explore the quadrature squeezing effect of the MSSVSs.

The coordinate operator is defined as $X=\left( a+a^{\dag }\right) /\sqrt{2}$%
\ and the momentum operator is defined as $P=\left( a-a^{\dag }\right) /(%
\sqrt{2}i)$. Their respective variances can expressed as follows \cite%
{21-1,21-2}
\begin{eqnarray}
\Delta ^{2}X &=&\left\langle a^{\dag }a\right\rangle -\left\vert
\left\langle a^{\dag }\right\rangle \right\vert ^{2}+\text{Re}(\left\langle
a^{\dag 2}\right\rangle -\left\langle a^{\dag }\right\rangle ^{2})+\frac{1}{2%
},  \notag \\
\Delta ^{2}P &=&\left\langle a^{\dag }a\right\rangle -\left\vert
\left\langle a^{\dag }\right\rangle \right\vert ^{2}-\text{Re}(\left\langle
a^{\dag 2}\right\rangle -\left\langle a^{\dag }\right\rangle ^{2})+\frac{1}{2%
}.  \label{b1}
\end{eqnarray}%
Their uncertainty relation obeys $\Delta ^{2}X\Delta ^{2}P\geq 1/4$. In
particular, a coherent (or vacuum) state holds $\Delta ^{2}X=\Delta
^{2}P=1/2 $. Generally, a quantum state is called squeezing if $\Delta
^{2}X<1/2$ or $\Delta ^{2}P$ $<1/2$.

In order to study the squeezing effect, one can firstly calculate the
general expected value $\left\langle a^{\dag k}a^{l}\right\rangle =\mathrm{Tr%
}\left( a^{\dag k}a^{l}\rho \right) $ with different integers $k,l$ for
quantum state $\rho $. By choosing proper integers $k,l$, one can obtain any
expected values one needed. In the process of calculation, one can resort to
the techniques $a^{\dag k}=\frac{d^{k}}{df^{k}}e^{fa^{\dag }}|_{f=0}$ and $%
a^{l}=\frac{d^{l}}{dg^{l}}e^{ga}|_{g=0}$.

For example, the SVS has%
\begin{equation}
\left\langle a^{\dag k}a^{l}\right\rangle _{\rho _{SV}}=\frac{d^{k+l}}{%
df^{k}dg^{l}}e^{\frac{1}{1-\lambda ^{2}}\frac{\lambda }{2}\left(
f^{2}+g^{2}\right) +\frac{\lambda ^{2}}{1-\lambda ^{2}}fg}|_{f=g=0}.
\label{b5}
\end{equation}%
When $k=l=1$, it leads to $\left\langle a^{\dag }a\right\rangle _{\rho
_{SV}}=\lambda ^{2}/\left( 1-\lambda ^{2}\right) $.$\allowbreak $ If $%
k=1,l=0 $ and $k=2,l=0$, we have $\left\langle a^{\dag }\right\rangle =0$
and $\left\langle a^{\dag 2}\right\rangle =\lambda /\left( 1-\lambda
^{2}\right) , $ respectively. Thus for the SVS $\rho _{SV}$, $\Delta ^{2}X=%
\frac{1}{2}e^{2r}$ and $\Delta ^{2}P=\frac{1}{2}e^{-2r}$. So the SVS has
squeezing effect for any nonzero $r$. While for the MSSVSs, we have%
\begin{eqnarray}
\left\langle a^{\dag k}a^{l}\right\rangle &=&\frac{1}{m!p_{d}\sqrt{\epsilon
_{4}}}\frac{d^{2m+k+l}}{d\mu ^{m}d\nu ^{m}df^{k}dg^{l}}  \notag \\
&&e^{\frac{\epsilon _{3}\epsilon _{7}}{\epsilon _{4}}\left( \mu
f+\allowbreak g\nu \right) +\allowbreak \frac{\epsilon _{3}\epsilon _{8}}{%
\epsilon _{4}}\left( \nu f+g\allowbreak \mu \right) }  \notag \\
&&\times e^{(\lambda ^{2}\tau _{1}+\frac{\epsilon _{5}}{\epsilon _{4}})fg+(%
\frac{1}{2}\lambda \tau _{1}+\frac{\epsilon _{6}}{\epsilon _{4}})\left(
f^{2}+g^{2}\right) }  \notag \\
&&\times e^{(1-\frac{\epsilon _{1}}{\epsilon _{4}})\mu \nu +\frac{\epsilon
_{2}}{\epsilon _{4}}\left( \mu ^{2}+\nu ^{2}\right) }|_{f=g=\mu =\nu =0}
\label{b3}
\end{eqnarray}%
with $\epsilon _{3}=\lambda \tau _{3}\sqrt{1-\eta _{2}}$, $\epsilon
_{5}=4\lambda \epsilon _{2}\epsilon _{3}^{2}-\left( 1+\lambda ^{2}\right)
\epsilon _{1}\epsilon _{3}^{2}$, $\epsilon _{6}=\left( 1+\allowbreak \lambda
^{2}\right) \epsilon _{2}\epsilon _{3}^{2}-\lambda \epsilon _{1}\epsilon
_{3}^{2}$, $\epsilon _{7}=\lambda \epsilon _{1}-2\epsilon _{2}$, and $%
\epsilon _{8}=\epsilon _{1}-2\lambda \epsilon _{2}$.
\begin{figure}[tbp]
\label{FigSQE} \centering\includegraphics[width=1.0\columnwidth]{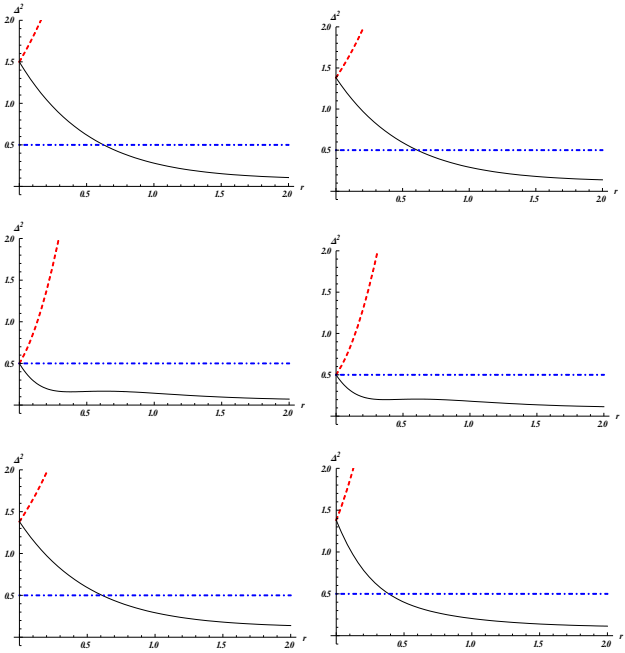}
\caption{(Colour online) Quadrature squeezing effect vesus $r$ for the MPSSV
with fixed $T=0.9$ and (row 1) $m=1$, (row 2) $m=2$, (row 3) $m=3$ and (left
column) $\protect\eta _{1}=\protect\eta _{1}=0$, (right column) $\protect%
\eta _{1}=\protect\eta _{1}=0.1$. Notice that the black solid line, the red
dashed line, and the blue dotdashed line are corresponding to $\Delta ^{2}P$%
, $\Delta ^{2}X$\ and $1/2$, respectively. Notice the threshold value is
different in each sub-figure.}
\end{figure}

In Fig.3, we plot the variation of $\Delta ^{2}P$ and $\Delta ^{2}X$ versus $%
r $ for several different cases at fixed $T=0.9$. It is clearly seen that
the variance of $\Delta ^{2}P$ monotonously decreases as $r$ increases for a
given $m$ and there exists squeezing in $P$ quadrature component within a
certain range of parameter $r$. For the case of $m=1$, only when the
squeezing parameter $r$ is bigger than a threshold value $r_{c}$, depending
on the loss factor, the MSSVS may have the possibility of the squeezing
effect (Noticing $r_{c}=0.626381$\ for $\eta _{1}=\eta _{2}=0$\ and $%
r_{c}=0.609918$\ for $\eta _{1}=\eta _{2}=0.1$). For the case of $m=2$, the
MSSVS always presents the squeezing effect for any nonzero $r$. For the case
of $m=3$, only when the squeezing parameter $r$ is bigger than a threshold
value $r_{c}$, the MSSVS may have the possibility of the squeezing effect,
where $r_{c}=0.396049$\ for $\eta _{1}=\eta _{2}=0$\ and $r_{c}=0.387008$\
for $\eta _{1}=\eta _{2}=0.1$, respectively.

\section{Photon number distribution}

Photon number distribution (PND) is defined by $P\left( n\right)
=\left\langle n\right\vert \rho \left\vert n\right\rangle $, which means the
probability of detecting $n$ photons in the field $\rho $. Using the
technique $\left\langle n\right\vert =\frac{1}{\sqrt{n!}}\frac{d^{n}}{ds^{n}}%
\left\langle 0\right\vert e^{sa}|_{s=0}$\ and $\left\vert n\right\rangle =%
\frac{1}{\sqrt{n!}}\frac{d^{n}}{dt^{n}}e^{ta^{\dag }}\left\vert
0\right\rangle |_{t=0}$, one can easily obtain the PNDs for the given
optical field $\rho $.

For the SVS $\rho _{SV}$, we have%
\begin{eqnarray}
P_{\rho _{SV}}\left( n\right) &=&\frac{\left( 1-\lambda ^{2}\right) ^{1/2}}{%
n!}\frac{d^{2n}}{ds^{n}dt^{n}}e^{\frac{\lambda }{2}(s^{2}+t^{2})}|_{s=t=0}
\notag \\
&=&\left\{
\begin{array}{cc}
\frac{n!\lambda ^{n}\left( 1-\lambda ^{2}\right) ^{1/2}}{2^{n}\left( \frac{n%
}{2}\right) !\left( \frac{n}{2}\right) !}, & n\text{ is even,} \\
0, & n\text{ is odd,}%
\end{array}%
\right.  \label{c5}
\end{eqnarray}%
which implies that the SVS contains only even-photon components, as seen
from Fig.4. This is one of the key characteristics of the SVS.

But what will happen to the PNDs for the MSSVSs? Noticing Eq.(\ref{a17}), we
obtain the PND of the MSSVSs expressed as%
\begin{eqnarray}
P\left( n\right) &=&\frac{1}{n!m!p_{d}\sqrt{\epsilon _{4}\kappa _{4}}}\frac{%
d^{2m+2n}}{d\mu ^{m}d\nu ^{m}ds^{n}dt^{n}}  \notag \\
&&e^{[1-\frac{\epsilon _{1}}{\epsilon _{4}}+\frac{\kappa _{3}^{2}}{\kappa
_{4}}\left( \kappa _{1}\kappa _{5}+4\kappa _{2}\kappa _{6}\right) ]\mu \nu
+(1-\frac{\kappa _{1}}{\kappa _{4}})st}  \notag \\
&&\times e^{[\frac{\epsilon _{2}}{\epsilon _{4}}-\frac{\kappa _{3}^{2}}{%
\kappa _{4}}\left( \kappa _{1}\kappa _{6}+\kappa _{2}\kappa _{5}\right)
]\left( \mu ^{2}+\allowbreak \nu ^{2}\right) +\frac{\kappa _{2}}{\kappa _{4}}%
\left( s^{2}+\allowbreak t^{2}\right) }  \notag \\
&&\times e^{\frac{\kappa _{3}}{\kappa _{4}}[\kappa _{7}\left( \mu t+s\nu
\right) +\kappa _{8}\left( \mu s+\nu t\right) ]}|_{\mu =\nu =s=t=0},
\label{c3}
\end{eqnarray}%
where $\kappa _{1}=1+\lambda ^{2}\tau _{1}+\frac{\allowbreak \epsilon _{5}}{%
\epsilon _{4}}$, $\kappa _{2}=\frac{1}{2}\lambda \tau _{1}+\frac{\epsilon
_{6}}{\epsilon _{4}}$, $\kappa _{3}=\frac{\epsilon _{3}}{\epsilon _{4}}$, $%
\kappa _{4}=\kappa _{1}^{2}-4\kappa _{2}^{2}$, $\kappa _{5}=8\lambda
\epsilon _{1}\epsilon _{2}-\left( 1+\allowbreak \lambda ^{2}\right) \left(
\epsilon _{1}^{2}+\allowbreak 4\epsilon _{2}^{2}\right) $, $\kappa
_{6}=\lambda \left( \epsilon _{1}^{2}+4\epsilon _{2}^{2}\right) -2\left(
1+\lambda ^{2}\right) \epsilon _{1}\epsilon _{2}$, $\kappa _{7}=\kappa
_{1}\epsilon _{7}-2\kappa _{2}\epsilon _{8}$, and $\kappa _{8}=\kappa
_{1}\epsilon _{8}-2\kappa _{2}\epsilon _{7}$.

In order to analyze the effect of loss on the PNDs of the MPSSVs, we depict
the PNDs in Fig.5 and Fig.6. From Fig.5 without loss ($\eta _{1}=0$, $\eta
_{2}=0$), we see that the MPSSVs contain only even-photon (odd-photon)
components if $m$ is even (odd), which agrees with the results of Ref.\cite%
{9-1}. However, we observe that, surprisingly, if there is the loss, the
MPSSVs contain all-photon components (see Fig.6). Moreover, the ratio
between even-component and odd-component can be adjusted by the value of $m$
and the loss factors $\eta _{1}$, $\eta _{2}$.
\begin{figure}[tbp]
\label{FigPND0} \centering\includegraphics[width=1.0\columnwidth]{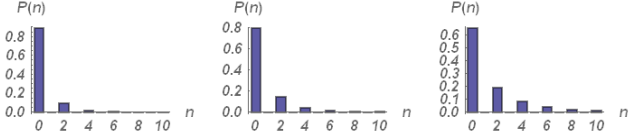}
\caption{(Colour online) PNDs for the SVs with $r=0.5$, $0.7$, $1$,
respectively.}
\end{figure}
\begin{figure}[tbp]
\label{FigPND1} \centering\includegraphics[width=1.0\columnwidth]{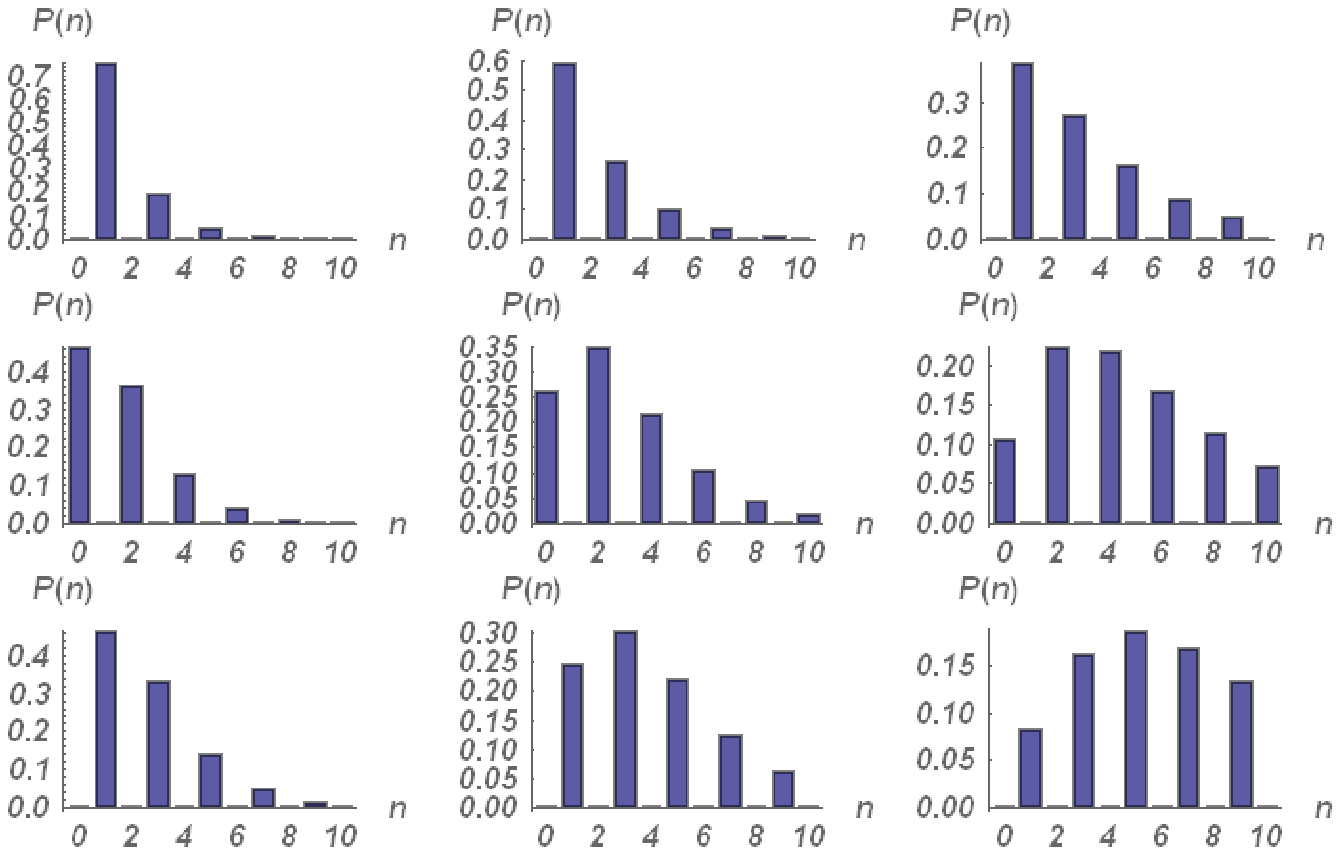}
\caption{(Colour online) PNDs for the MPSSVs wihtout loss ($\protect\eta %
_{1}=0$, $\protect\eta _{2}=0$) with given other parameters. (Row 1) with $%
T=0.9$, $m=1$; (Row 2) with $T=0.9$, $m=2$; (Row 3) with $T=0.9$, $m=3$.
Column 1, column 2, and column 3 are corresponding to $r=0.5$, $r=0.7$, and $%
r=1$, respectively.}
\end{figure}
\begin{figure}[tbp]
\label{FigPND2} \centering\includegraphics[width=1.0\columnwidth]{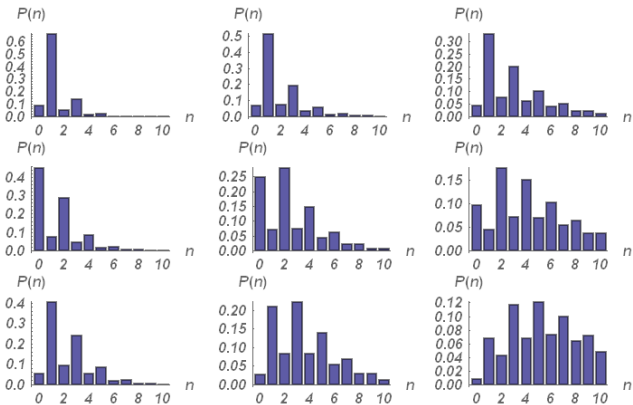}
\caption{(Colour online) PNDs for the MPSSVs wiht loss ($\protect\eta %
_{1}=0.1$, $\protect\eta _{2}=0.1$) with given other parameters. (Row 1)
with $T=0.9$, $m=1$ and different $r$; (Row 2) with $T=0.9$, $m=2$ and
different $r$; (Row 3) with $T=0.9$, $m=3$ and different $r$. Column 1,
column 2, and column 3 are corresponding to $r=0.5$, $r=0.7$, and $r=1$,
respectively.}
\end{figure}

\section{Wigner function}

Wigner function $W\left( \beta \right) $ for quantum state $\rho $ can be
defined by\cite{22}%
\begin{equation}
W\left( \beta \right) =\frac{2}{\pi }\mathrm{Tr}[\rho D\left( \beta \right)
\left( -1\right) ^{a^{\dag }a}D^{\dag }\left( \beta \right) ],  \label{d1}
\end{equation}%
where $D\left( \beta \right) =\exp \left( \beta a^{\dag }-\beta ^{\ast
}a\right) $ is the usual displacement operator with $\beta =\left(
x+iy\right) /\sqrt{2}$.

For the SVS, we know%
\begin{equation}
W_{\rho _{SV}}\left( \beta \right) =\frac{2}{\pi }e^{-\frac{2(1+\lambda ^{2})%
}{1-\lambda ^{2}}\left\vert \beta \right\vert ^{2}+\allowbreak \frac{%
2\lambda }{1-\lambda ^{2}}\left( \beta ^{2}+\beta ^{\ast }{}^{2}\allowbreak
\right) },  \label{d3}
\end{equation}%
which is Gaussian and implies that the SVS are Gaussian states (see Fig.7).
While for the MSSVSs $\allowbreak $, we have%
\begin{eqnarray}
W\left( \beta \right) &=&\frac{1}{\pi m!p_{d}\sqrt{\epsilon _{4}\kappa _{9}}}%
e^{-(\frac{\kappa _{1}}{\kappa _{9}}-\frac{1}{2\kappa _{9}})\left\vert \beta
\right\vert ^{2}+\frac{\kappa _{2}}{\kappa _{9}}\left( \beta
^{2}+\allowbreak \beta ^{\ast }{}^{2}\right) }  \notag \\
&&\frac{d^{2m}}{d\mu ^{m}d\nu ^{m}}e^{[1-\frac{\epsilon _{1}}{\epsilon _{4}}+%
\frac{\kappa _{3}^{2}}{\kappa _{9}}(4\kappa _{2}\kappa _{6}+\kappa
_{1}\kappa _{5}-\frac{1}{2}\kappa _{5})]\mu \nu }  \notag \\
&&\times e^{[\frac{\epsilon _{2}}{\epsilon _{4}}+\frac{\kappa _{3}^{2}}{%
\kappa _{9}}(\frac{1}{2}\kappa _{6}-\kappa _{1}\kappa _{6}-\kappa _{2}\kappa
_{5})]\left( \mu ^{2}+\nu ^{2}\right) }  \notag \\
&&\times e^{[(\frac{\kappa _{1}}{\kappa _{9}}-\frac{1}{2\kappa _{9}})\kappa
_{3}\epsilon _{8}-\frac{\allowbreak 2\kappa _{2}}{\kappa _{9}}\kappa
_{3}\epsilon _{7}]\left( \nu \beta +\mu \beta ^{\ast }\right) }  \notag \\
&&\times e^{[(\frac{\kappa _{1}}{\kappa _{9}}-\frac{1}{2\kappa _{9}})\kappa
_{3}\epsilon _{7}-\frac{\allowbreak 2\kappa _{2}}{\kappa _{9}}\kappa
_{3}\epsilon _{8}]\left( \mu \beta +\nu \beta ^{\ast }\right) }|_{\mu =\nu
=0},  \label{d2}
\end{eqnarray}%
where $\kappa _{9}=(\kappa _{1}-\frac{1}{2})^{2}-4\kappa _{2}^{2}$.
\begin{figure}[tbp]
\label{FigWF0} \centering\includegraphics[width=1.0\columnwidth]{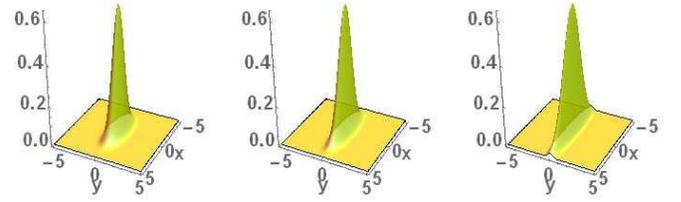}
\caption{(Colour online) Wigner functions for the SVs with $r=0.5$, $0.7$, $%
1 $, respectively.}
\end{figure}
\begin{figure}[tbp]
\label{FigWF1} \centering\includegraphics[width=1.0\columnwidth]{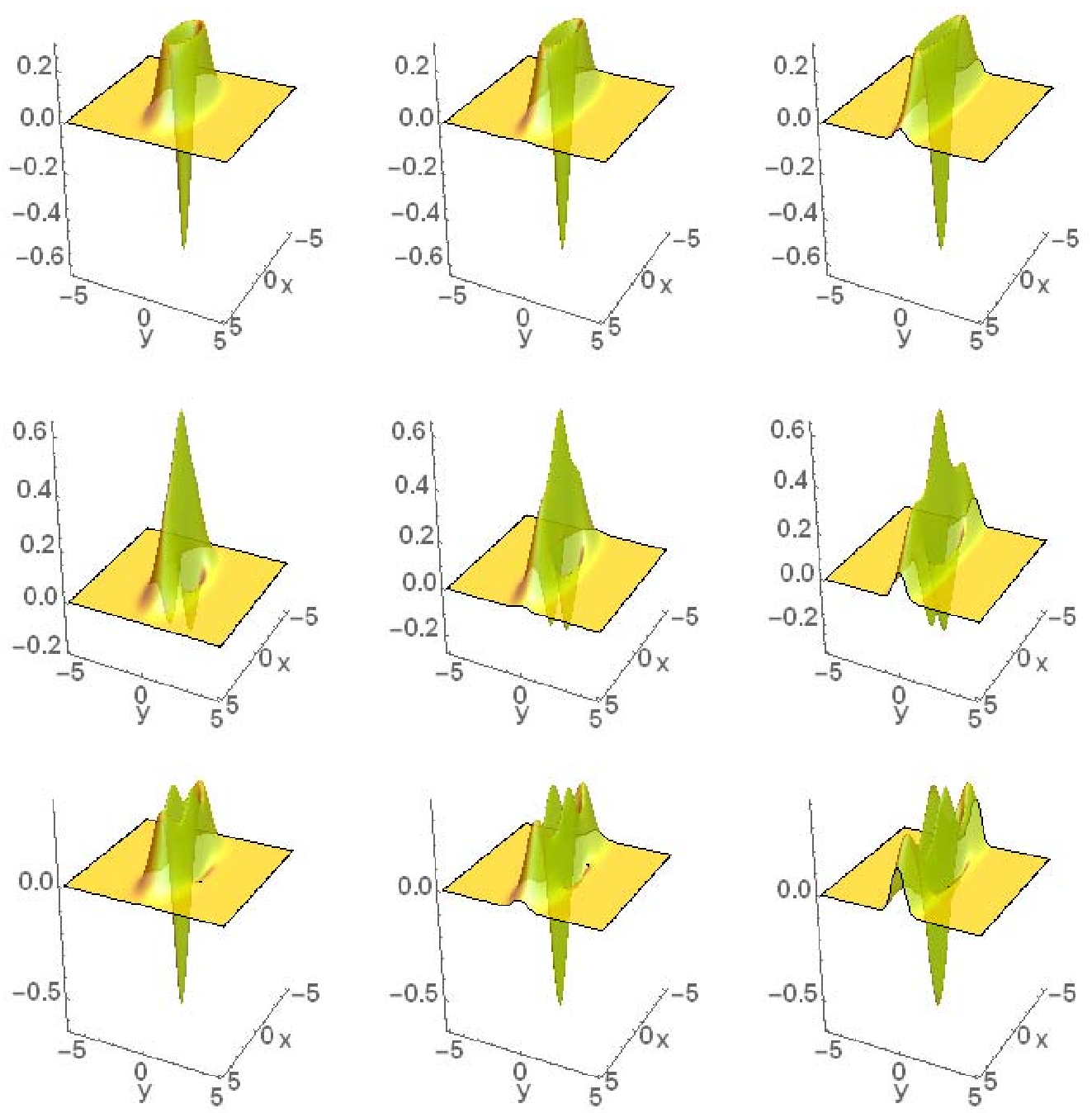}
\caption{(Colour online) Wigner functions for the MPSSV without loss ($%
\protect\eta _{1}=0$, $\protect\eta _{2}=0$) with given other parameters.
(Row 1) with $T=0.9$, $m=1$; (Row 2) with $T=0.9$, $m=2$; (Row 3) with $%
T=0.9 $, $m=3$. Column 1, Column 2, and Column 3 are corresponding to $r=0.5$%
, $r=0.7$, and $r=1$, respectively.}
\end{figure}
\begin{figure}[tbp]
\label{FigWF2} \centering\includegraphics[width=1.0\columnwidth]{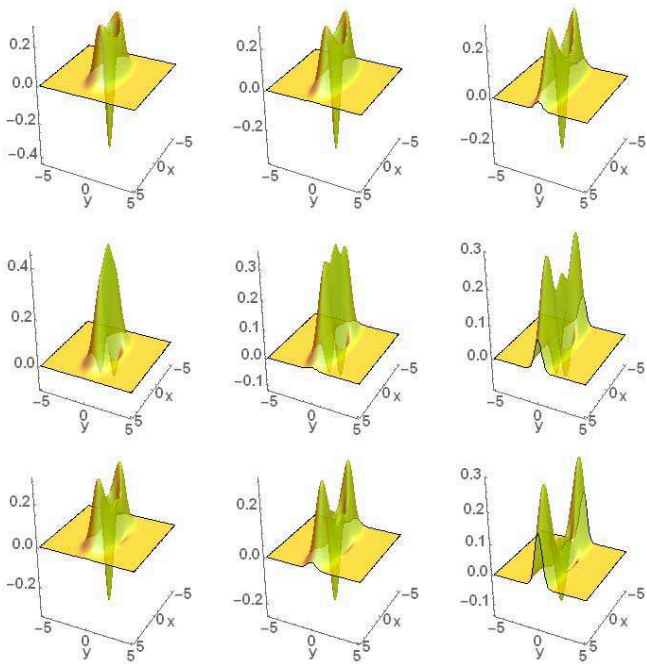}
\caption{(Colour online) Wigner functions for the MPSSV without loss ($%
\protect\eta _{1}=0.1$, $\protect\eta _{2}=0.1$) with given other
parameters. (Row 1) with $T=0.9$, $m=1$; (Row 2) with $T=0.9$, $m=2$; (Row
3) with $T=0.9$, $m=3$. Column 1, Column 2, and Column 3 are corresponding
to $r=0.5$, $r=0.7$, and $r=1$, respectively.}
\end{figure}

According to Eq.(\ref{d2}), we plot the Wigner functions of the MSSVSs in
Fig.8 without loss and in Fig.9 with loss in phase space. Clearly, the
Wigner functions of the MSSVSs is non-Gaussian in phase space. The surfaces
without loss in Fig.8 are smoother than those with loss in Fig.9. As an
evidence of the nonclassicality of the state\cite{23}, there are some
negative regions of the Wigner function in phase space (see Figs. 8 and 9).
Moreover, the distribution of Wigner function can reflect the
non-Gaussianity of quantum states\cite{24,25}.

\section{Conclusion}

To summarize, we present a conditional scheme of generating the MSSVSs in
the presence of pure-loss channels. By adjusting the relative interaction
parameters (including $r$, $\eta _{1}$, $\eta _{2}$, $T$ and $m$), a broad
class of MSSVSs with figure of merit can be obtained. For the theoretical
model, we have given the complete description of density operator of the
optical fields in terms of CF. Analytical derivation and numerical
simulation for the properties of the MSSVSs are explored in detail. Compared
with the original SVS and the MSSVSs without loss, some interesting results
effected by the loss are summarized as follows: (1) The losses change a
threshold value of original squeezing parameter, which may present the
appearance of squeezing effect; (2) The losses will let the MSSVSs contain
all-photon components (including odd-photon and even-photon); (3) The losses
make the distribution of the Wigner function more complex.

\begin{acknowledgments}
This project was supported by the National Natural Science Foundation of
China (No.11665013).
\end{acknowledgments}

\end{document}